\documentclass[aps,prx,twocolumn,showpacs,superscriptaddress,floatfix]{revtex4-1}

\usepackage{amsmath}
\usepackage{amssymb}
\usepackage{graphicx}
\usepackage{grffile} 
\usepackage{float}
\usepackage{color}
\usepackage{subfigure}
\usepackage[normalem]{ulem}
\usepackage{hyperref}
\usepackage{lipsum,calc}
\usepackage[shortlabels]{enumitem}
\hypersetup{
    colorlinks=true, 
    citecolor=blue,  
    urlcolor=blue,
    linkcolor=blue
}

\newcommand{\half}{\frac{1}{2}}
\newcommand{\kk}{\mathbf{k}}
\newcommand{\qq}{\mathbf{q}}

\newcommand{\A}{\mathbf{A}}

\newcommand{\LK}[1]{\textcolor{black}{#1}}

\newcommand{\trel}{t_\mathrm{rel}}
\newcommand{\tave}{t_\mathrm{ave}}

\newcommand{\imsigma}{\mathrm{Im}\ \Sigma}
\newcommand{\tdelay}{t_\mathrm{delay}}

\begin{document}

\author{A.~F.~Kemper}
\email{akemper@ncsu.edu}
\affiliation{Department of Physics, North Carolina State University, Raleigh, NC27695}

\author{O.~Abdurazakov}
\affiliation{Department of Physics, North Carolina State University, Raleigh, NC27695}

%
%
\author{J.~K.~Freericks}
\affiliation{Department of Physics, Georgetown University, Washington, DC 20057, USA}
%

\title{General principles for the non-equilibrium relaxation of populations in quantum materials}

\begin{abstract}
We examine the problem of how excited populations of electrons relax after they have been excited by a pump. We include three of the most important relaxation processes: (i) impurity scattering; (ii) Coulomb scattering; and (iii) electron-phonon scattering. The relaxation of an excited population of electrons is one of the most fundamental processes measured in pump/probe experiments, but its interpretation
remains under debate.  We show how several common assumptions about non-equilibrium relaxation that are pervasive in the field
may not hold under quite general conditions. The analysis shows that non-equilibrium relaxation is more complex than previously thought, but it yields to recently developed theoretical methods in non-equilibrium theory. 
In this work, we show how one can use many-body theory to properly interpret and analyze these complex systems.
We focus much of the discussion on implications of these results for experiment.
\end{abstract}

\maketitle

\section{Introduction}

In the analysis of non-equilibrium theory and experiments, it is common to lean on intuition developed from equilibrium physics and from linear response, even in circumstances where the system is driven far from equilibrium.
The experiments are typically performed with a pump/probe setup, where the pump drives the system out of equilibrium and out of the domain where our equilibrium intuition applies. The measurement is made with a wide variety of probes in a similar variety of contexts,
and they have been the subject of rapid development.
Theory has also improved beyond Boltzmann equation and effective temperature approaches,
with exact methods being developed in one dimension (via density matrix renormalization group\cite{schollwock_dmrg_2005}) and in infinite dimensions (via non-equilibrium dynamical mean-field theory\cite{freericks_nonequilibrium_2006,aoki_nonequilibrium_2014}).
While much work has been accomplished in both cases, \LK{ there are a number of assumptions that underlie a significant fraction
of the analysis in the field which do not always hold.}
We describe four of these assumptions in detail here and fully consider three of them, leaving the resolution of the fourth to later work. 

The four \LK{assumptions} are as follows: (i) many-body systems must relax after excitation; 
(ii) the self-energy governs the relaxation rate; (iii) the time domain allows one to separate the relaxation rates from different scattering processes; and (iv) when electrons are coupled to phonons, they rapidly scatter amongst themselves to create hot thermal electrons that subsequently relax with the phonons until they both reach a common final temperature.

The reasons why each of these \LK{assumptions may be readily violated} follows from a rather straightforward analysis:

(i) Isolated electronic systems cannot relax back to the state before the pump (equilibrium) because their total energy (given by their initial energy plus the energy imparted by the pump) is conserved. Hence, while they can rearrange the energy amongst their constituents, in many cases this does not evolve to a thermal distribution. Recent work on the eigenstate thermalization hypothesis says that systems without additional conserved quantum numbers do evolve to a thermal distribution, while those with additional conserved quantities will evolve to generalized Gibbs distributions which maintain the additional conservation laws.\cite{rigol_thermalization_2008}
In addition, pumped systems sometimes are excited into metastable states or novel non-equilibrium phases, which may relax in
unknown ways (and not to the pre-pump state), because these phases are not present in thermal equilibrium.\cite{stojchevska_ultrafast_2014}
We will show that pure impurity scattering does not cause relaxation to thermal equilibrium while pure Coulomb scattering should. When electrons are attached to a phonon bath with electron-phonon coupling, then they can relax in a way that removes energy from the electronic system and eventually returns the system to the temperature of the phonon bath.

(ii) When the many-body system does relax, early work in many-body physics suggested that it is the imaginary part of the self-energy that governs the relaxation rate
\cite{allen_theory_1987,spicka_long_2005,*spicka_long_2005-1,*spicka_long_2005-2,kemper_mapping_2013,sentef_examining_2013},
similar to the situation in equilibrium\cite{galitskii1958translation,Mahan}.
While one can show this holds for inelastic scattering processes
in non-equilibrium when both the coupling and pump are weak,
it is violated by stronger interactions \LK{and higher pump fluence}\cite{kemper_effect_2014}. An even more significant violation has been reported
in experiments\cite{gierz_non-equilibrium_2014,yang_inequivalence_2015} but a full explanation of this effect does not exist
yet. In this work, we will point towards a resolution of this conundrum.

(iii) Matthiessen's rule\cite{matthiessen_rule_1864} governs the relaxation of most systems in linear response. It says that one {\it adds} the relaxation rates for different scattering processes to yield {\it one net relaxation rate} (similar to how resistors add as reciprocals when in a parallel circuit). As with the electrical circuit  analogy, one should only see the net relaxation rate in the time dynamics. Yet, in numerous experiments one does see a separation of relaxation rates for excited populations. We will show this naturally occurs in non-equilibrium relaxation and leads to violations of Matthiessen's rule. 

(iv) Finally, the hot-electron model, which has been employed in many different contexts, can be easily shown to be incorrect. An equation of motion for the population shows that whenever one has a fluctuation-dissipation-like relationship, 
($A^<(\omega) \propto f(\omega) \mathrm{Im}A^R(\omega)$), where the lesser quantities (propagators
and self-energies) are given 
by the imaginary parts of retarded quantities multiplied by some distribution function (it need not even be an equilibrium distribution), then the population no longer relaxes with time! In fact, one can easily show that it is the {\it deviations from the hot-electron model} that govern the non-equilibrium relaxation\cite{kemper_relationship_16,kemper_role_16}. We will summarize the argument for why this must occur.
Given the wide application of hot-electron models to analyzing pump/probe experiments, this is our most important result. We discuss later under what circumstances hot-electron model analyses make sense (as good approximations) and under what situations they fail.

\LK{Quite generally, these assumptions break down due to the fundamental difference between measurement time and quasiparticle lifetime.
By nature, equilibrium measurements (through linewidth, for example) observe the quasiparticle lifetime that arises from Fourier
transformation along relative time; these observations access the dephasing/decay of the correlation functions. On the other hand,
non-equilibrium measurements see the energy transfer between various subsystems rather than this dephasing.}

\LK{Some of these assumptions have been previously treated theoretically within Boltzmann equation approaches\cite{rethfeld_ultrafast_2002,baranov_theory_2014}, whose authors find similar results regarding the limitations of the hot electron
model and the separation of time scales.  However, Boltzmann equation approaches by nature do not capture the
emergence of the history kernel, and thus provide no description of the retarded nature of the interactions and frequency-dependence
of the electron Green's functions.  Nevertheless, these works find discrepancies between the inherent dephasing time of
a Fermi gas and the population dynamics after the pump\cite{rethfeld_ultrafast_2002},
and the critical appearance of bottlenecks\cite{baranov_theory_2014}, which we further elucidate here.}

To demonstrate that these assumptions \LK{may be readily violated for simple systems,}
we compare and contrast relaxation from impurities, coupling of electrons to phonons, electron-electron scattering, and various combinations.
\LK{Although these are relatively simple scattering processes that are well-understood in equilibrium, we will show that their
characteristics in the time domain are not simple extensions from equilibrium.  Furthermore, we will use them as counter examples
to show that even in simple cases the assumptions may be violated.}
We will show how some of the examples behave differently because they cannot fully relax, and how their temporal dynamics shows clear violations of Matthiessen's rule, with the origin arising from details for how energy is transferred from electrons to phonons. We also describe the implications of this work for the interpretation of experiments, in particular, emphasizing the different characteristic behaviors of the different scattering processes.

\section{Model and method}

We will investigate the aforementioned scattering channels: electron-impurity, electron-phonon, and electron-electron (Coulomb) 
scattering
through the Hubbard-Holstein model with local interactions
\begin{align}
\mathcal H= &\sum_{\kk,\sigma} \epsilon(\kk) c^\dagger_{\kk,\sigma} c^{\phantom\dagger}_{\kk,\sigma} 
		+ \bar U \sum_i c^\dagger_{i\uparrow} c^{\phantom\dagger}_{i\uparrow} c^\dagger_{i\downarrow} c^{\phantom\dagger}_{i\downarrow}
		+ \sum_i \bar V_i c^\dagger_{i,\sigma} c^{\phantom\dagger}_{i,\sigma}\nonumber\\
		&+ \Omega  \sum_i b_i^\dagger b^{\phantom\dagger}_i  
		+ g \sum_{\kk,\qq,\sigma} c_{\kk+\qq,\sigma}^\dagger c^{\phantom\dagger}_{\kk,\sigma} 
		\left( b^{\phantom\dagger}_\qq + b_{-\qq}^\dagger \right) 
\end{align}
where the individual terms represent the kinetic energy of electrons with a dispersion $\epsilon(\kk)$, the energy of Einstein phonons with a frequency $\Omega$, and a coupling between them whose strength is given by $g$.  Here, $c^\dagger_\alpha (c_\alpha)$ are the standard creation (annihilation) operators for an electron in state $\alpha$; similarly, $b^\dagger_\alpha (b_\alpha)$ creates (annihilates) a phonon in state $\alpha$.  The $\bar U$ and $\bar V_i$ terms represent the on-site Coulomb interaction and random impurity scattering, respectively.
We study a square lattice dispersion with nearest and next-nearest neighbor hopping ($t_{nn}$ and $t_{nnn}$),
\begin{align}
\epsilon(\kk) =& -2 t_{nn} \left[ \cos(k_x) + \cos(k_y)\right] + \nonumber \\
& 4t_{nnn} \cos(k_x) \cos(k_y) -\mu
\end{align}
where $\mu$ is the chemical potential.  We have used the convention that $\hbar=c=e=1$, which makes inverse energy the unit of time.  We choose $t_{nn}=0.25$ eV, $t_{nnn} = 0.075$ eV, and $\mu=-0.255$ eV, and use an inverse temperature $\beta=100$/eV.

\begin{figure}[t]
	\includegraphics[clip=true, trim=0 0 0 -40,width=0.99\columnwidth]{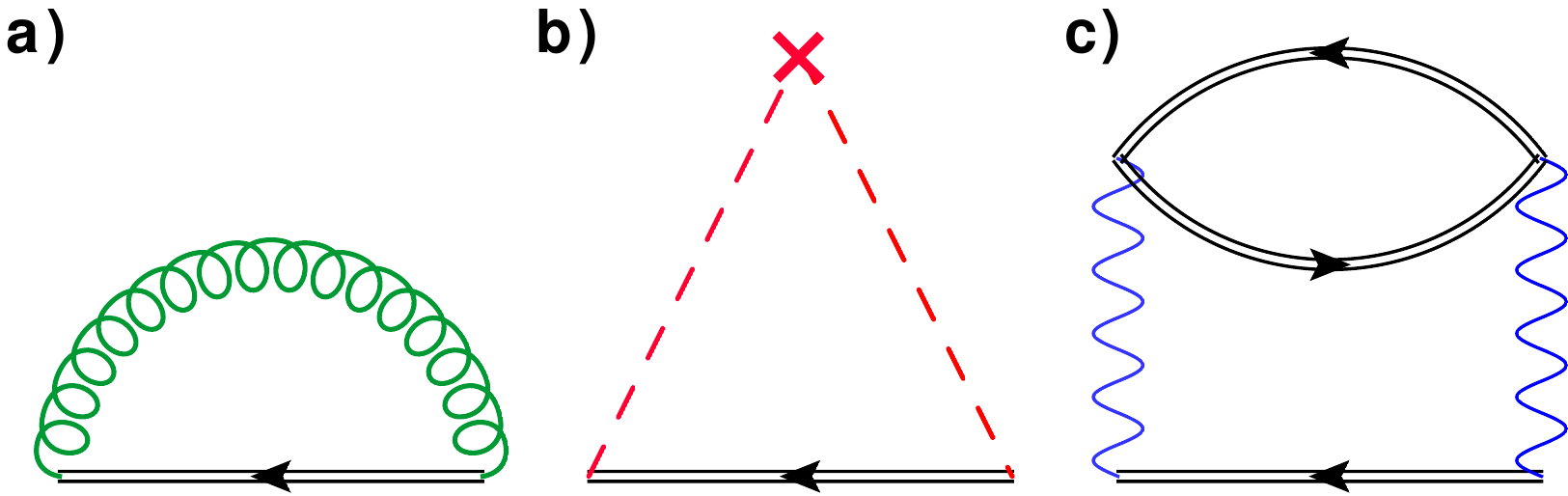}
	\caption{(Color online) Self-energies used in this study. 
	a) Electron-phonon self-energy in the Migdal approximation.
	b) Impurity self-energy in the Born approximation.
	c) Second order electron-electron self-energy.
	All of these are treated self-consistently.
	}
	\label{fig:selfenergies}
\end{figure}

The self-energies for the three different scattering mechanisms are shown in the Feynman diagrams in Fig.~\ref{fig:selfenergies}.
The electron-phonon interaction is treated at the self-consistent Born level (Fig.~\ref{fig:selfenergies}a),
with the self-energy given by
\begin{align}
\Sigma_{\mathrm{el-ph}}^\mathcal{C}(t,t') = i g^2 \  G_\mathrm{loc}^\mathcal{C}(t,t')\ D^\mathcal{C}_0(t,t').
\end{align}
Here, $D^\mathcal{C}_0(t,t')$ is the non-interacting phonon propagator\cite{Mahan}, and
$G^\mathcal{C}_\mathrm{loc}(t,t') = N_\kk^{-1} \sum_\kk \bar G^\mathcal{C}_\kk(t,t')$ 
is the local Green's function.  The superscript $\mathcal{C}$ denotes
that the quantity lives on the two-time Keldysh contour\cite{keldysh_1965}.  
In this formulation, we assume the phonons have an infinite heat capacity, so their temperature does not rise due to coupling with the excited electrons. Hence, they maintain the same phonon propagator for all times. This approximation is valid for situations where the amount
of energy put into the system isn't large enough to cause a notable change in the phonon dynamics\cite{murakami_interaction_2015}.

Similarly, the impurity scattering self-energy after averaging (Fig.~\ref{fig:selfenergies}b) is given by
\begin{align}
\Sigma_{\mathrm{el-imp}}^\mathcal{C}(t,t') = V^2 \  G_\mathrm{loc}^\mathcal{C}(t,t').
\label{eq:sigma_imp}
\end{align}
$V$ is obtained from impurity averaging as $V^2 = n_\mathrm{imp} \bar V^2$, where $n_\mathrm{imp}$
is the density of impurities, and we have neglected the momentum dependence of the scattering
matrix element.

The electron-electron interactions are included at the level of self-consistent conserving second-order perturbation theory 
(Fig.~\ref{fig:selfenergies}c) via
\begin{align}
\Sigma_{\mathrm{el-el}}^\mathcal{C}(t,t') = \bar U^2 G_\mathrm{loc}^\mathcal{C}(t,t')  
G_\mathrm{loc}^\mathcal{C}(t,t') G_\mathrm{loc}^\mathcal{C}(t',t) .
\end{align}
which is equivalent to a dynamical mean-field approximation with a conserving perturbative impurity solver. This approximation neglects the momentum dependence of the two-particle bubble, which increases the phase space for electron-electron scattering, but should produce accurate results as long as the interaction is not too large.
For local interactions, the Hartree and Fock terms only shift the chemical potential.

\begin{figure}[b]
	\includegraphics[width=0.99\columnwidth]{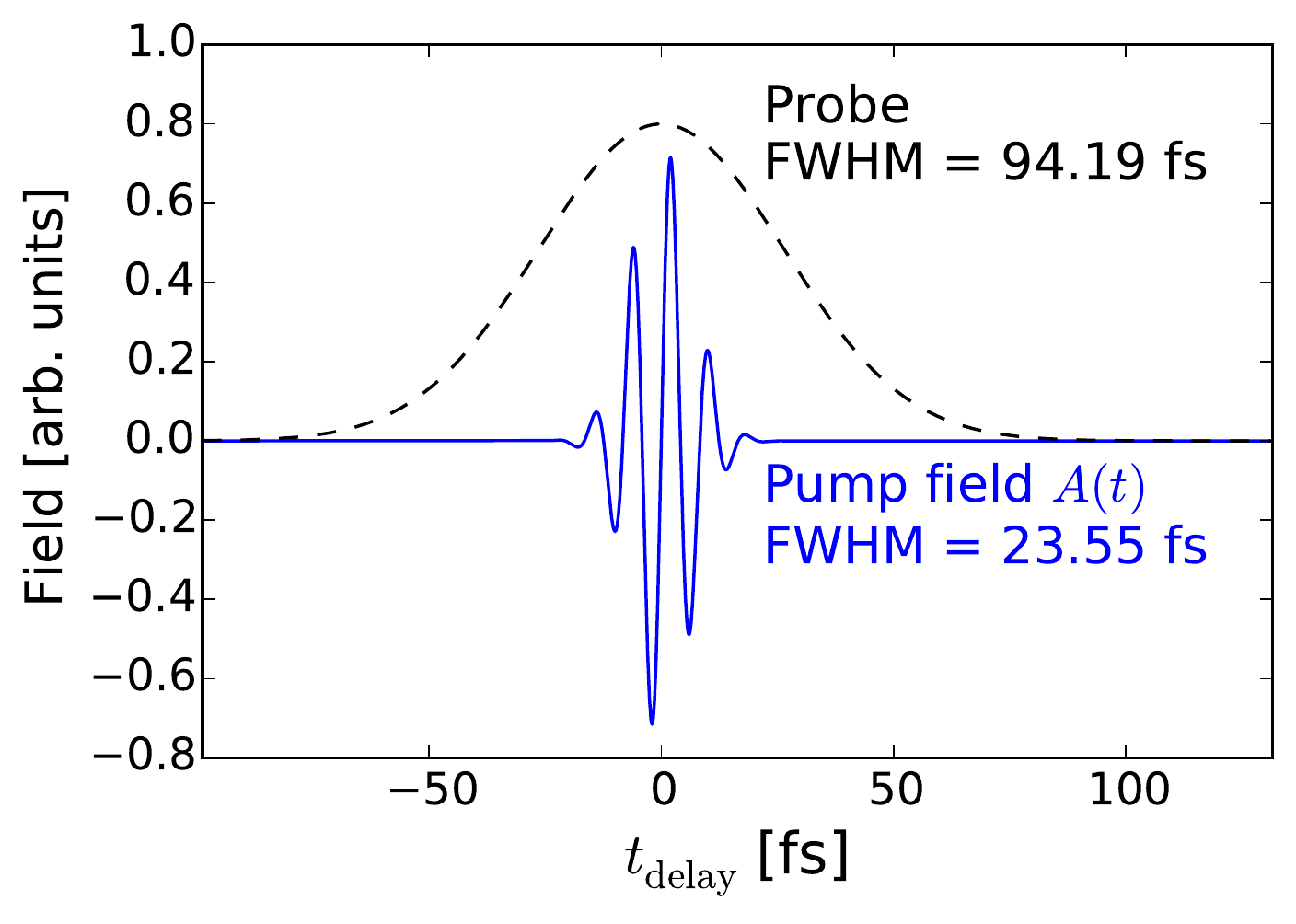}
	\caption{(Color online) Pump vector potential $A(t)$ and probe profiles used in this work. 
	The probe profile is the envelope function for the probe pulse.}
	\label{fig:field_profile}
\end{figure}

\begin{figure*}[ht]
	\includegraphics[width=0.9\textwidth]{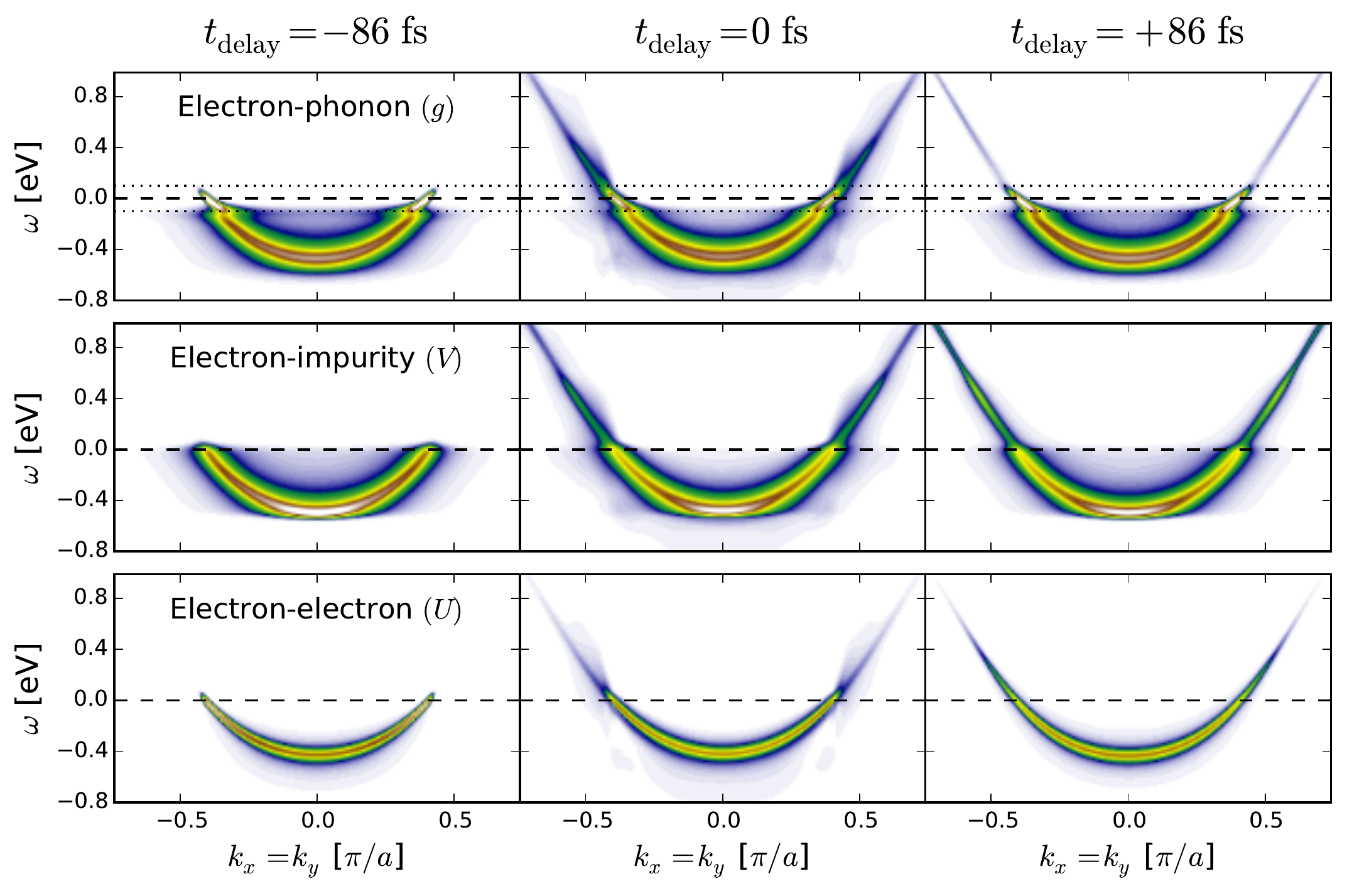}
	\caption{(Color online) False-color images of time-resolved single-particle spectral functions, at the start of (left), during (middle)
	and after (right) the pump for the individual electron-phonon, electron-impurity, and electron-electron interactions.
	Here, the interaction strengths are $g^2=V^2=U^2=0.02$ eV. The dotted line indicates the phonon frequency
	$\Omega=\pm0.1$ eV}
	\label{fig:trarpes}
\end{figure*}
Note that we did not include any mixed self-energy diagrams between the different scattering channels.  This is not a requirement,
but it allows for a clear differentiation between the different scattering mechanisms themselves, and it is the lowest-order contribution
in perturbation theory for each scattering channel.

In all three cases, there exists a sum rule for the interactions at this level of perturbation theory.
As was discussed previously\cite{freericks_nonequilibrium_2014,kemper_effect_2014}, the frequency-integrated electron-phonon interactions obey
a sum rule.  The impurity and Coulomb scattering self-energies also obey similar sum rules. They are as follows:
\begin{subequations}
\begin{align}
{\rm Im}\Sigma^R_{\mathrm{el-ph}}(t,t) =& -g^2[\langle x^2(t)\rangle-\langle x(t)\rangle^2]\\
=&-g^2 (2n_B(\Omega/T) + 1) \\
{\rm Im}\Sigma^R_{\mathrm{el-imp}}(t,t) =& -V^2 \\
{\rm Im}\Sigma^R_{\mathrm{el-el}}(t,t) =& - \bar U^2 n(1-n)\equiv-U^2
\label{eq:sumrules}
\end{align}
\end{subequations}
where $n=\langle n_\uparrow\rangle=\langle n_\downarrow\rangle$ is the electron density per spin and
$n_B(\Omega)$ is the Bose occupation of the phonon mode $\Omega$.
These identities are true at all times, and hold individually even when all
three types of interactions are present.
To put
the parameter strength for the various interactions on an equal footing, we define an interaction strength $U^2 = \bar U^2 n (1-n)$
which takes into account the dependence of the electron-electron scattering on the electron density.

The equations of motion for the Green's functions are solved on the contour by using
the methods described in Ref.~\onlinecite{stefanucci_book}.
The field is included via the Peierls substitution\cite{r_peierls_33} $\mathcal \kk(t) = \kk - \A(t)$, where the vector
potential $\A(t)$ is treated in the Hamiltonian gauge. We use a pump of the form 
$A(t) = A_\mathrm{max} \exp(-t^2/2\sigma^2)\sin(\omega t)$
in the diagonal $(11)$ direction 
with $\omega=0.5$ eV.
The field is illustrated in Fig.~\ref{fig:field_profile}. The full width at half max of the field is approximately 24~fs, 
while the width of the probe is approximately 94~fs (see below). \LK{The time scales for the fields are chosen for
computational reasons, as are the interaction strengths and related time scales.  However, since we are
focused on the post-pump dynamics, the time and energy scales in the pump do not have an imprint on the
physics discussed below; rather, the pump is only a mechanism to make an excitation.}

\subsection{Observables}
The single-particle angle-resolved photoemission spectral functions are obtained from the Green's functions via
\cite{freericks_constant_2016}
\begin{align}
&P(\kk,\omega,t_\mathrm{delay}) = \nonumber\\
&~~~~~\mathrm{Im} \int dt_1 dt_2 p(t_1,t_2,t_\mathrm{delay}) e^{i\omega (t_1-t_2)} G_{\bar\kk(t_1,t_2)}^<(t_1,t_2)
\end{align}
where $p(t_1,t_2,t_\mathrm{delay})$ is a two-dimensional normalized Gaussian with isotropic width $\sigma_p=\sigma$
centered at $t_1=t_2=\tdelay$:
\begin{align}
&p(t_1,t_2,t_\mathrm{delay})=\nonumber\\
&~~~~~\frac{1}{2\pi\sigma^2}\exp\left(-\frac{ (t_1-\tdelay)^2+(t_2-\tdelay)^2}{2\sigma^2}\right).
\end{align}
The Gaussian probe profile along $\tdelay$ is shown in Fig.~\ref{fig:field_profile}. 
The shift in momentum due to the Peierls substitution has to be corrected to
obtain gauge-invariant spectra through a time-dependent shift of the momentum\cite{bertoncini_jauho,v_turkowski_book} via
\begin{align}
\bar \kk(t,t') = \kk + \frac{1}{t-t'}\int_{t'}^t  \A(\bar t) d\bar t.
\end{align}

Below, we also consider the amount of excited electrons by evaluating the electron density above $k_F$:
\begin{align}
n(k>k_F) = -i \sum_{k>k_F} G^<_\kk(t,t'=t)
\label{eq:n_above_kf}.
\end{align}

\section{Results}

To set the stage for the discussion, we will outline the oft-used conceptual framework underlying the \LK{assumptions.}
First, in equilibrium, one typically identifies the quasiparticle lifetime with the self-energy through:
\begin{align}
\frac{1}{\tau(\nu)} = -2\mathrm{Im}\ \Sigma^R(\nu)
\end{align}
where we have combined the quasiparticle quantum numbers into a single index $\nu$. This relation was suggested
to hold in the time domain within certain limits with $\tau$ as the population decay rate rather than the single particle lifetime
\cite{allen_theory_1987,spicka_long_2005,*spicka_long_2005-1,*spicka_long_2005-2,kemper_mapping_2013,sentef_examining_2013},
but was experimentally shown not to hold in at least a few cases\cite{gierz_non-equilibrium_2014,yang_inequivalence_2015}.
One of the consequences of this line of thought is that, since the self-energy is always present, relaxation must always occur.

Next, in equilibrium the scattering rates for each of the
relevant processes encoded in the self-energy add linearly for weak coupling. Based on this concept
time-resolved ARPES experiments are commonly analyzed
by considering the contributions of various scattering processes to the population decay rate based on estimates
from equilibrium experiments.
Although the addition of the scattering rates should give rise to a single exponential decay rate, this is often not observed \textemdash\
a multi-exponential type of behavior is seen instead.  This \LK{agrees with} the mode of thinking that the processes in the
time domain should separate according to their fundamental timescales.  
Based on converting the characteristic energy
scales into time scales and ranking them,
the strong interaction (often electron-electron) is thought to occur first, followed by the
weaker interactions (e.g. electron-phonon scattering).

Below, we will show that these concepts are erroneous.  The population decay rate is not simply connected
to the self-energy, and the scattering rates do not always separate.  Rather, the population decay rate is set by a difference
between effective distributions of the populations and the self-energy, and any separation of scattering rates arises from energy
bottlenecks.

We begin by considering the effects of individual electron-phonon, electron-impurity, and electron-electron (Coulomb) scattering
in a pump-probe setup.
Figure~\ref{fig:trarpes} shows snapshots of the time-resolved ARPES spectra for the three individual interactions under consideration
(movies of the three processes are available in the supplemental materials).
To make a proper comparison the interaction strengths, as defined through the sum rules in Eqs.~\ref{eq:sumrules}, are kept
constant.
The equilibrium spectra at $t_\mathrm{delay}=-86$ fs show the typical hallmarks of a band of electrons interacting with phonons, impurities,
and internally through Coulomb scattering.  

The system with phonon scattering shows a kink in the spectra at the phonon frequency
together with a sharpening of the spectra within the phonon window (defined as $|\omega| < \Omega$).  
\LK{This window plays a key role determining the population dynamics at low energies due to phase space restrictions
for phonon emission.\cite{sentef_examining_13,kemper_effect_2014,rameau_energy_2016}}

The system with impurity scattering
has a more of less homogenous scattering rate proportional to the density of states (DOS). 
Coulomb scattering, within second order perturbation theory,
gives rise to a self-energy (and thus a scattering rate) that is given by $\imsigma \sim -(\omega^2 + \pi^2T^2)$.  The temperature
here is sufficiently low that the $\omega^2$ term dominates, which is reflected in the spectra in the increase in the linewidth at high energies.

The spectra at $t_\mathrm{delay}=0$ fs are at the point where the pump and probe overlap.  In all three cases, there is excited spectral
weight above the Fermi level;  there are some qualitative distinctions between the spectra that originate from the particular
interactions (e.g. the weakening of the kink in the phonon-scattering case), but the excitation is present in all cases.
\LK{This is also similar to what one would envision from the hot-electron approach. But, if the electronic system fully relaxes to a hot temperature, we will show below that it will never subsequently relax to a lower temperature.}

\subsection{tr-ARPES for individual scattering channels}

\begin{figure}
	\includegraphics[width=\columnwidth]{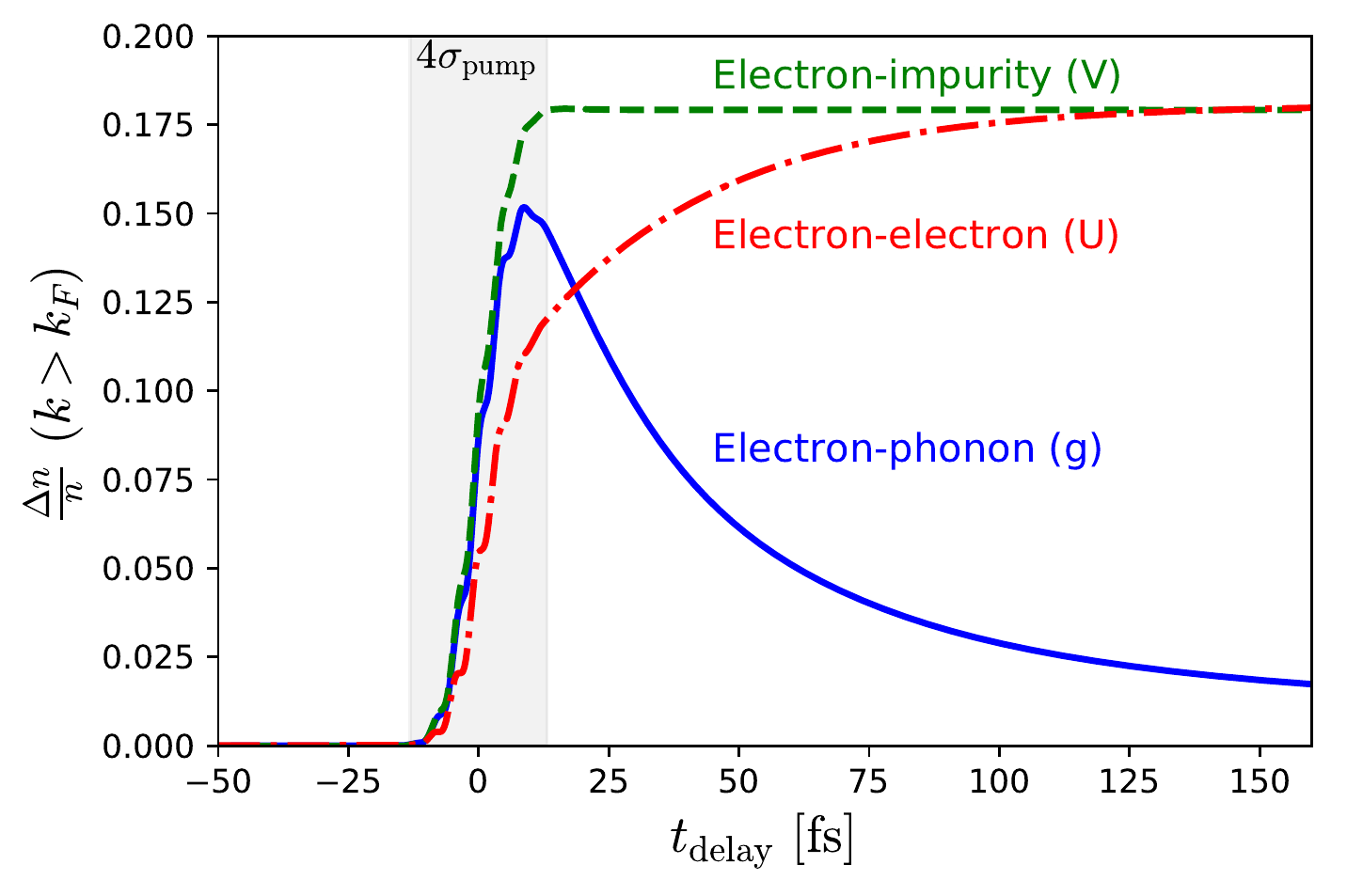}
	\caption{(Color online) Quasiparticle weight above the Fermi level for the individual electron-phonon, electron-impurity, and electron-electron
	interactions. The shaded region indicates $\pm 2\sigma$ of the pump field.}
	\label{fig:unocc_pop}
\end{figure}

The final snapshot at $t_\mathrm{delay}=+86$ fs. is taken long after the pump.  At this point, the system with phonon scattering has
partially completed its return to equilibrium, as has been studied in some detail
previously\cite{kemper_mapping_2013, sentef_examining_2013,kemper_effect_2014}.  This is clearly reflected in the partial disappearance
of the excited spectral weight above the Fermi
level (see also Fig.~\ref{fig:unocc_pop}).  The systems where only impurity or Coulomb scattering are present are in
stark contrast to the former \textemdash\ after the pump, essentially they essentially do not return to the equilibrium state before the pump.
However, although these two share the characteristic of a new steady state after the pump, the appearance of the spectra
is quite different.  For impurity scattering, there is essentially no change in the spectra immediately after the pump,
which agrees with previous analytic work\cite{kemper_relationship_2016} showing a steady state once momentum equilibriation
is achieved. This is expected because the final state should be a generalized Gibbs\cite{rigol_thermalization_2008} state which will not look thermal.
On the other hand, 
the system with Coulomb scattering exhibits some short-time dynamics and quickly settles into the state seen in the figure. 
The short time dynamics are caused by internal thermalization of the electrons, and the steady state is an
elevated temperature state, as expected by the eigenstate thermalization hypothesis.

Fig.~\ref{fig:unocc_pop} illustrates the difference between the three scattering processes through the excited population above
the Fermi level (see Eq.~\ref{eq:n_above_kf}).
The electron-phonon interactions behave as expected, with an excitation followed by a return to equilibrium.
On the other hand, the impurity and Coulomb scattering cases end up in a state with a sizable persistent increase in the population above
the Fermi level, although the Coulomb scattering system takes some time to get there as it scatters electrons from below $E_F$.

These observations immediately demonstrate \LK{a violation of two of the assumptions}: many-body systems do not always relax to the equilibrium state before the pump even long after the pulse ends
[assumption (i)], nor does the self-energy directly govern the relaxation rate [assumption (ii)].

\subsection{Impurity and Phonon scattering}
\begin{figure}[t]
	\includegraphics[width=0.99\columnwidth]{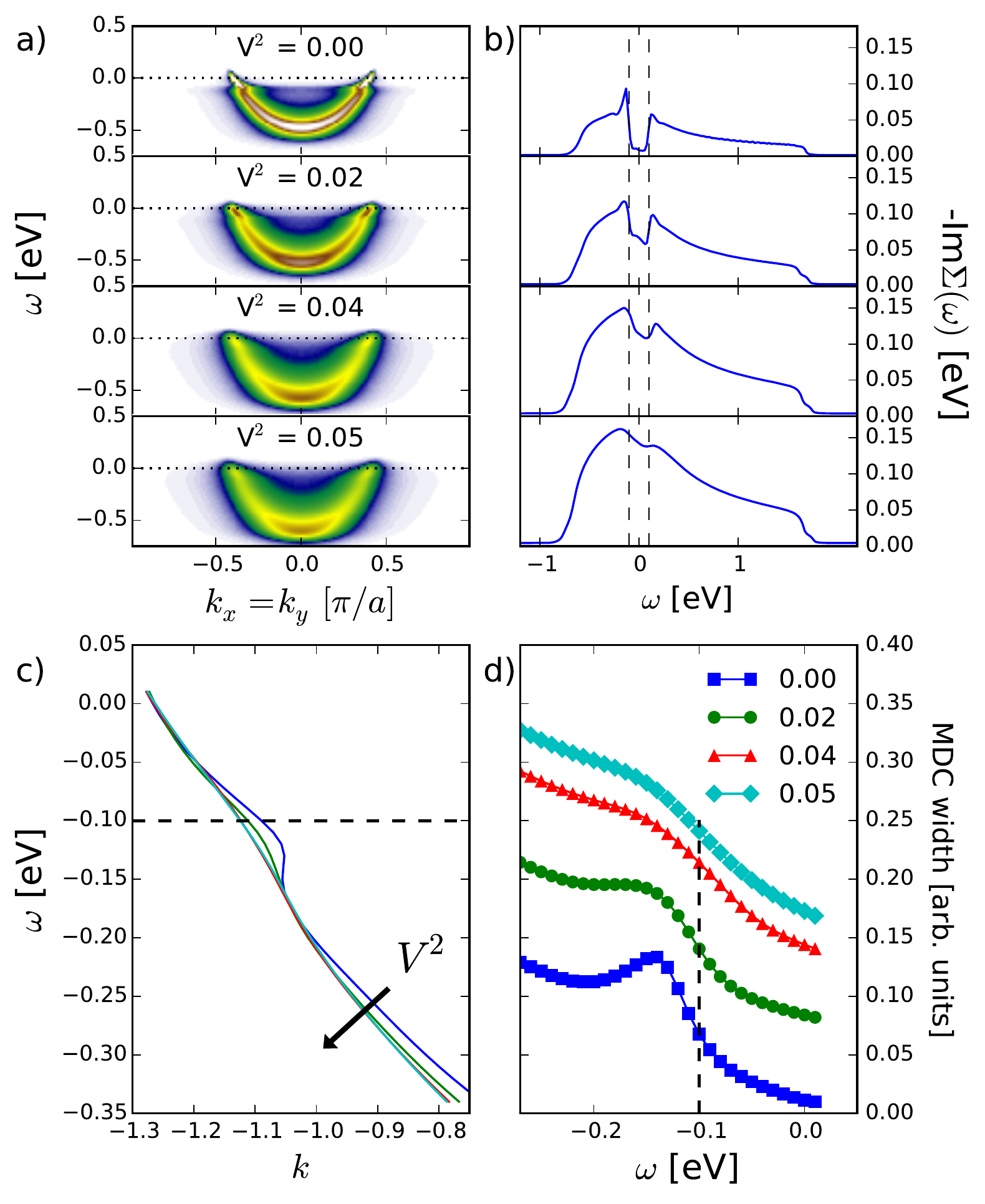}
	\caption{(Color online) Effect of combined electron-impurity and electron-phonon scattering in equilibrium. 
	a,b) Equilibrium ARPES spectra and self-energies $(-\imsigma(\omega))$ for various impurity scattering strengths $V$.
	The phonon scattering strength is kept constant at $g^2=0.02$ eV. c,d) Extracted dispersion $\epsilon(\kk)$ and MDC widths
	from the ARPES spectra.}
	\label{fig:imp_equilibrium}
\end{figure}

Figures~\ref{fig:trarpes} and \ref{fig:unocc_pop} show that the different interactions exhibit qualitatively different temporal dynamics,
beyond the quantitative aspect that is expected due to their different nature. 
Here, we will show the error in \LK{assumption} (ii): \textit{the self-energy governs the relaxation rate},
by considering the
combination of impurity and phonon scattering in some detail.  Fig.~\ref{fig:imp_equilibrium} shows the equilibrium effect of increasing impurity scattering on top of a constant phonon scattering strength .
The impurity scattering self-energy is proportional to the density of states,
which adds (according to Matthiessen's rule) to the existing phonon \LK{self-energy}, increasing the overall single-particle scattering rate and
filling in the phonon window. This is reflected in the spectra, in which the phonon kink can be seen when $V^2=0$, but it is
rapidly obscured as impurity scattering increases.  This can be further analyzed by fitting the spectral momentum distribution cuts (MDCs), which yield
the dispersion $\epsilon(\kk)$ and MDC line widths.  The dispersion shows the kink gradually disappearing, and the MDC
widths show a monotonic increase as well as the disappearance of the step in the single-particle scattering rate. At the
highest impurity scattering strength, where $V^2 = 2.5\times g^2$, the phonon window and kink are effectively gone.

\begin{figure}[htpb]
	\includegraphics[width=0.9\columnwidth, clip=true, trim= 0 30 0 30]{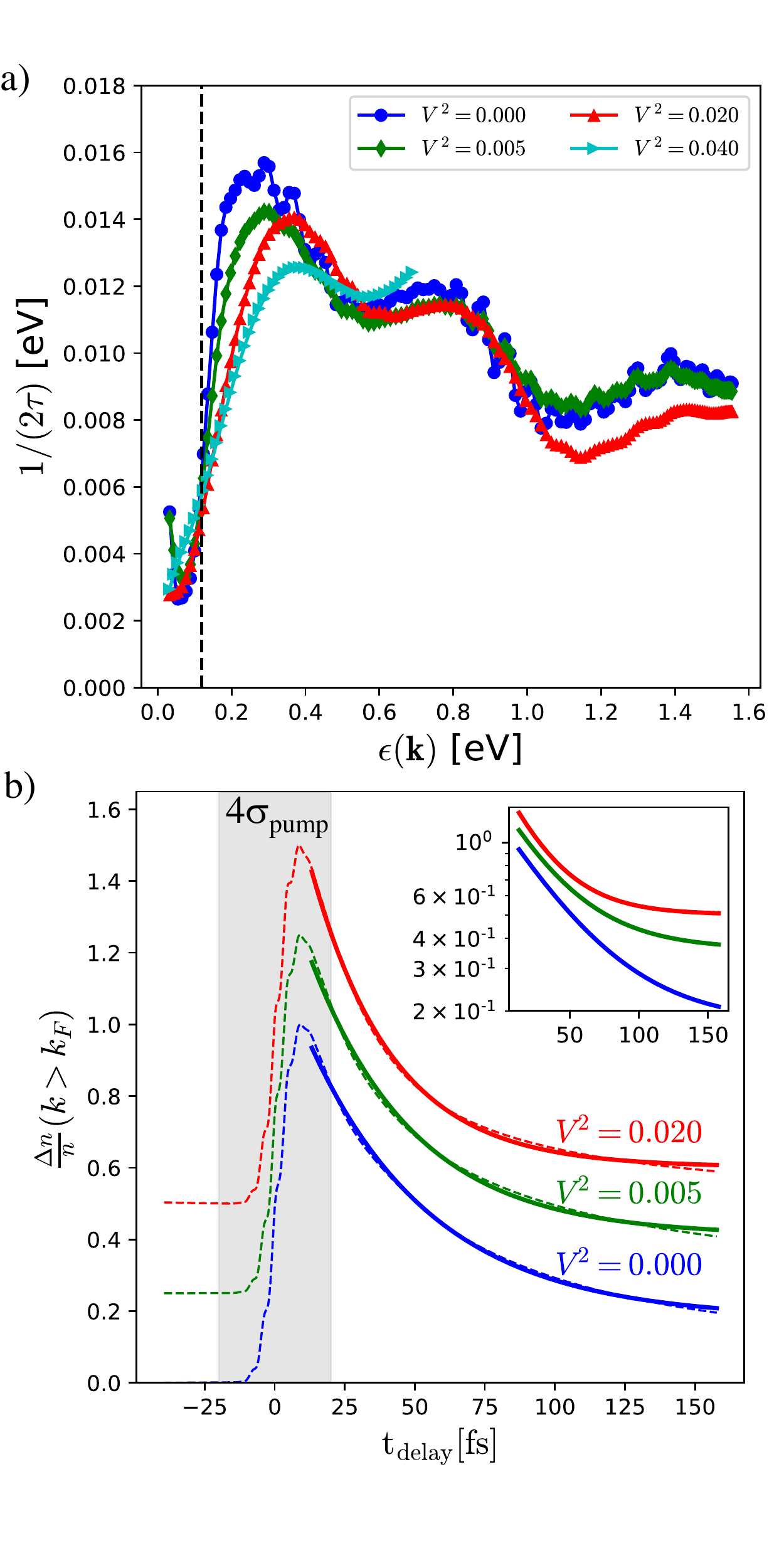}
	\caption{(Color online) 
	a) Population decay rates after excitation showing a weak dependence on the impurity scattering rate $V$ (values for $V^2$ are in eV$^2$). 
	b) Normalized quasiparticle weight above the Fermi level for phonon and impurity scattering (offset for clarity).
	The solid lines are \textit{single} exponential fits to the region where the solid line is shown, which should not work
	if timescales separate.
	The shaded region indicates $\pm 2\sigma$ of the pump field.
	Inset: Semilogarithmic plot of the same data.
	}
	\label{fig:imprates}
\end{figure}

\textit{If the self-energy were to control the relaxation rate of the population then the relaxation rate should increase as the impurity scattering
increased.}  Fig.~\ref{fig:imprates}a) plots the decay rates as a function of quasiparticle energy $\epsilon(\kk)$, as the impurity
scattering is increased from $V^2=0$.  The figure clearly shows that the dramatic effect on the equilibrium measures (MDC line width,
self-energy) are wholly absent in the non-equilibrium population decay rates.
This demonstrates \LK{a violation of assumption} (ii): \textit{the self-energy governs the relaxation rate} by contradiction;
there are substantial changes in the self-energy due to the presence and strength of the impurity scattering 
that are \underline{not} observed in the non-equilibrium population dynamics where the changes are minor.
As we will see, aside from the weak-coupling limit of pure electron-phonon scattering\cite{sentef_examining_2013}, 
this result turns out to be the generic behavior of nonequilibrium systems with a range of different scattering mechanisms.

For this mixed-interaction case,
it appears that the electron-phonon scattering is the dominant mechanism in the return to equilibrium.  We have previously
argued that this is because phonons can draw energy out of the system\cite{rameau_energy_2016}. Indeed, an isolated electronic system must 
conserve its total energy and hence it cannot dissipate its excess energy.  Impurities scatter
quasiparticles throughout the Brillouin zone
but do so while conserving energy
and thus
cannot drive the system back to equilibrium with a lower temperature. In this case, \LK{assumption} (iii), which says Matthiessen's rule is violated, does hold, but in an overstated way \textemdash\ since the impurities are inefficient in relaxing the electrons, we see much too mild of a dependence of the scattering rate on the impurity scattering strength. Hence, we do not see a separation of the scattering into an impurity-based scattering and an electron-phonon based scattering.  

This is further highlighted in Fig.~\ref{fig:imprates}b), which shows the population above the Fermi level evaluated
according to Eq.~\ref{eq:n_above_kf}.  \LK{If assumption (iii) were to hold, then there should be a double exponential decay,
with a time constant for each process.  To illustrate that this is not the case, we show that the curves can be reasonably fit
using a {\it single} exponential (although some deviations are visible). Thus, the time scales clearly do not separate simply based on
the presence of multiple scattering channels with different coupling constants.}

\begin{figure}[htpb]
	\includegraphics[width=0.99\columnwidth]{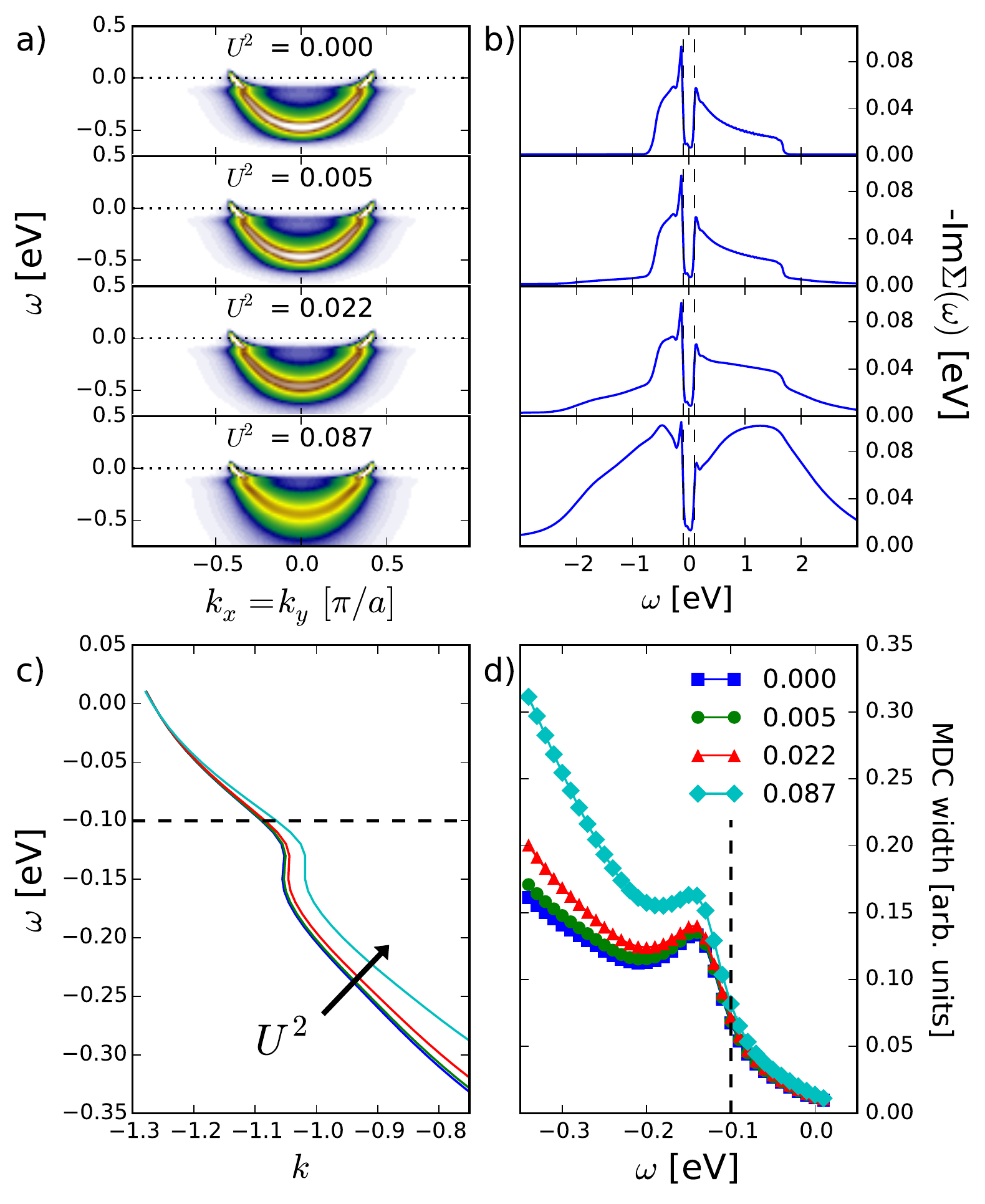}
	\caption{(Color online) Effect of combined electron-electron and electron-phonon scattering in equilibrium. 
	a,b) Equilibrium ARPES spectra and self-energies $(-\imsigma(\omega))$ for various Coulomb scattering strengths $U$.
	The phonon scattering strength is kept constant at $g^2=0.02$ eV. c,d) Extracted dispersion $\epsilon(\kk)$ and MDC widths
	from the ARPES spectra.}
	\label{fig:ee_equilibrium}
\end{figure}

\subsection{Coulomb and Phonon scattering}
\label{sect:ee_and_ep}

Coulomb, or electron-electron scattering, is critically different from impurity scattering in that it inelastic; it can redistribute energy among
quasiparticles (although the total energy in the quasiparticle system is conserved).  This can be seen in Fig.~\ref{fig:trarpes}
as a difference between the spectra at $\tdelay=0$ fs and $\tdelay=86$ fs.  Shortly after the pump, the electrons re-arrange
to form a steady state that is different from equilibrium (although it is thermal).
When phonon scattering is also present, a complex interplay between the two scattering channels emerges
which we will use to demonstrate \LK{a violation of assumption}
(iii) \textit{the time domain allows one to separate the relaxation rates from different scattering processes.}
In addition, \LK{violation of assumption} (ii) will be illustrated once again.

We consider the dynamics of a system subject to both
Coulomb and phonon scattering.  Fig.~\ref{fig:ee_equilibrium} shows the same analysis as was performed for the impurity scattering:
the effects of increasing Coulomb scattering on the equilibrium spectra.  Here, since the Coulomb scattering self-energy is proportional
to $\omega^2$ at low temperatures, the main effect on the self-energy occurs at high energies and the phonon window
remains intact (Fig.~\ref{fig:ee_equilibrium}b).  This is reflected in the spectra (Fig.~\ref{fig:ee_equilibrium}a),
which continue to show the phonon kink, but gain an increased line width at higher
energies.  This is further confirmed quantitatively through the MDC fits in both the dispersion and line width analysis 
(Figs.~\ref{fig:ee_equilibrium}c, \ref{fig:ee_equilibrium}d)

\begin{figure}[h]
	\includegraphics[width=0.9\columnwidth, clip=true, trim= 0 40 0 20]{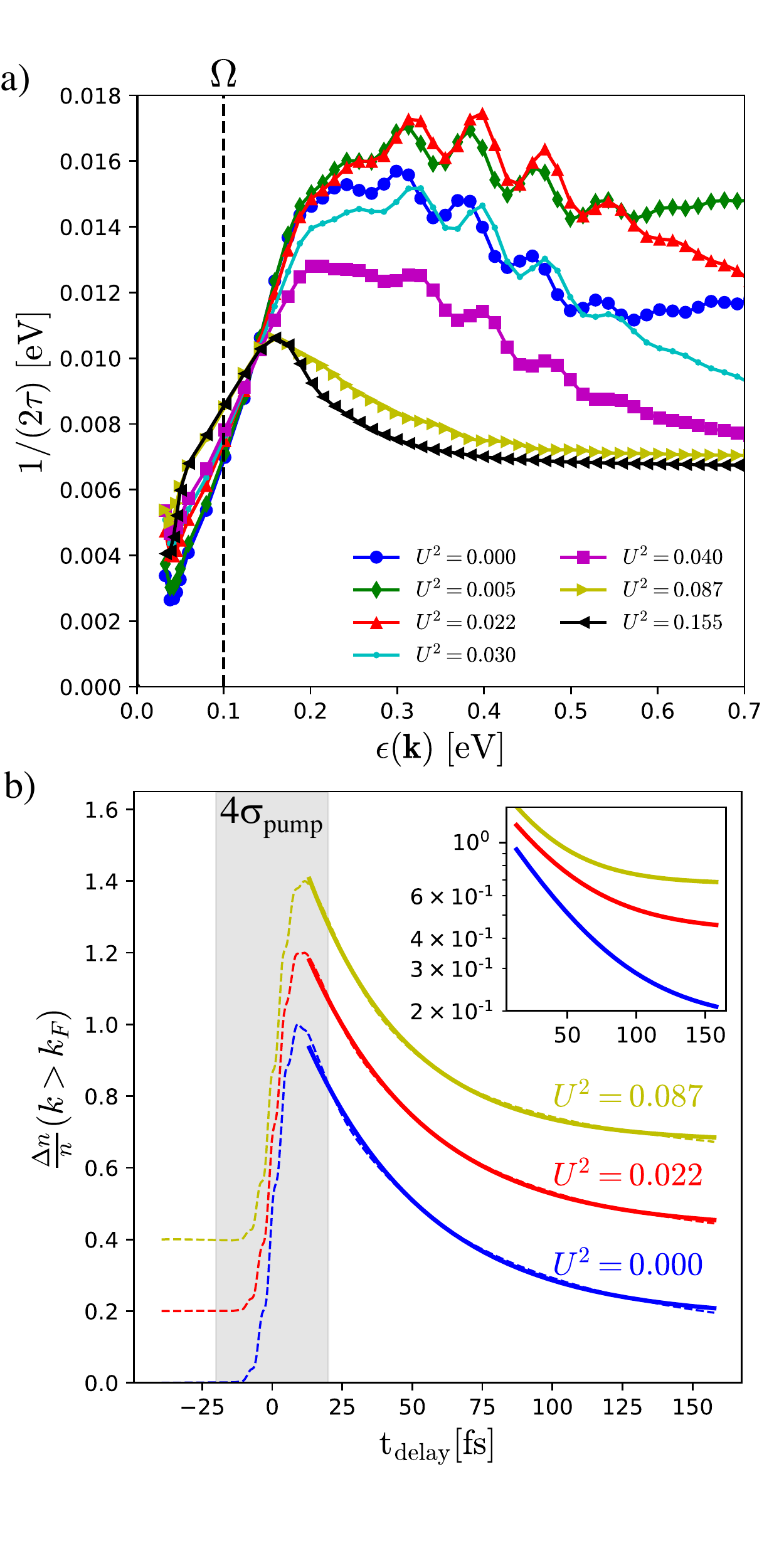}
	\caption{(Color online) a) Population decay rates after excitation showing a non-monotonic dependence on the Coulomb scattering strength $U$
	(values of $U^2$ are in eV$^2$.) \LK{Outside the phonon window}, after an initial increase, the decay rates decrease. Some of the data are reproduced from 
	Ref.~\cite{rameau_energy_2016} 
	b) Quasiparticle weight above the Fermi level for phonon and Coulomb scattering (offset for clarity).
	The solid lines are \textit{single} exponential fits to the region where the solid line is shown, which should not work
	if timescales separate.
	The shaded region indicates $\pm 2\sigma$ of the pump field.
	Inset: Semilogarithmic plot of the same data.
	}
	\label{fig:coulombrates}
\end{figure}

Turning now to the population dynamics in the time domain, we observe a very different situation from impurity scattering
as is clear from Figure~\ref{fig:coulombrates}a.  There is a notable effect with increasing Coulomb scattering, but the effect is not the
linear increase one would expect from Matthiessen's rule.
While this \textit{is} observed in the
self-energies, which increase with the addition of Coulomb scattering (c.f. Fig.~\ref{fig:ee_equilibrium}b), 
the population decay rates initially show a modest increase
which subsequently reverses with a stronger Coulomb scattering strength, in particular at high energies
above the phonon window.  The phonon window itself remains essentially visible, although the size of the step in $1/2\tau$
decreases.  
We can clearly see a violation of Matthiessen's rule, and that the self-energy does not directly govern the relaxation rate.
Furthermore, \LK{just as for the case of combined phonon and impurity scattering}, the population decay in time (Fig.~\ref{fig:coulombrates}b) does \textit{not} show two distinct relaxation times \LK{with a double
exponential form}, nor is there a
marked increase in the decay rate with increasing $U$.
It is dominated by the electron-phonon relaxation time, with a small increase in the rate due to Coulomb rearrangement
of the quasiparticles enabling slightly more efficient phonon emission.
The decay is controlled by the phonons because they provide the effective path, and thus a bottleneck of sorts, for the
electrons to lose their excess energy.

\section{Discussion}

\subsection{What does Mathiessen's rule suggest?}

Is it instructive to reiterate the conclusions one would draw from a simple Mathiessen's rule analysis.  If the decay rates are
to be proportional to the excited density to arrive at an exponential decay, we may expect the populations to obey a simple differential equation:
\begin{align}
\frac{dn}{dt} = -\frac{1}{\tau_1} n -\frac{1}{\tau_2}n + \cdots = -\frac{1}{\tau_\mathrm{eff}} n.
\end{align}
Regardless of the number of channels, one would arrive at a single exponential decay rate.  Moving beyond that, to arrive at
a double exponential (with decay constants $\gamma_1$ and $\gamma_2$, 
one would need a \textit{second order} differential equation of the form
\begin{align}
\frac{d^n}{dt^2} = - \left(\gamma_1 + \gamma_2\right) \frac{dn}{dt} - \gamma_1 \gamma_2 n
\end{align}

The exact equations of motion may in limited cases be reduced to the former, but generally 
do not fit either of these\cite{kemper_relationship_16}.
Hence, one should not expect these simple behaviors to occur. In no circumstances is there any simple explanation for multiple relaxation times in terms of the equations of motion for the population.

\subsection{Origin of the breakdown of the assumptions}

The contrast between Figures~\ref{fig:imp_equilibrium} and \ref{fig:ee_equilibrium} (equilibrium) and 
Figures~\ref{fig:trarpes},~\ref{fig:unocc_pop},~\ref{fig:imprates}, and \ref{fig:coulombrates} (non-equilibrium) are the central results that demonstrate
\LK{violation of the assumptions:}

\begin{enumerate}
\item Assumption (i) \textit{many-body states relax after excitation} is contradicted by the simple examples
of single-channel scattering through either impurities or Coulomb scattering 
neither
of which decay but both of which are interacting many-body systems.  
%
\item Assumption (ii) \textit{the self-energy governs the relaxation
rate} is contradicted by a comparison between the equilibrium self-energy (Figs.~\ref{fig:imp_equilibrium} and \ref{fig:ee_equilibrium})
and the population dynamics 
(Figs.~\ref{fig:trarpes},~\ref{fig:unocc_pop}, \ref{fig:imprates}, \ref{fig:coulombrates}). 
The self-energies monotonically increase as
the second scattering channel is added, whereas the population decay rates show entirely different behavior. This explains experimental results which show large differences between equilibrium linewidths and non-equilibrium decay rates.\cite{gierz_non-equilibrium_2014,yang_inequivalence_2015}.
\item Assumption (iii) \textit{the time domain allows one to separate the relaxation rates from different scattering processes}
is contradicted by Figs.~\ref{fig:imprates} and \ref{fig:coulombrates}.  \LK{As evidenced from Fig.~\ref{fig:imprates}, }impurity scattering
(in the Born approximation) can be removed entirely from the consideration
of energy-resolved decay rates since it
plays no role in the return to equilibrium.  Electron-electron
scattering plays a more intricate role when it is added to phonon scattering (c.f. Fig.~\ref{fig:coulombrates}), but it does not appear that electron-electron scattering always
happens first, followed by phonon scattering, even though the energy scales are quite different.
In either case, going to the time domain does not separate out the time-scales for scattering.
In fact, in situations where Matthiessen's rule holds there is only one effective relaxation time determined by the different
scattering mechanisms. 
In situations where Matthiessen's rule is violated, and the different timescales do emerge in non-equilbrium, 
separation of timescales occurs due to other reasons such as bottlenecks associated with the transfer of energy from the electrons 
to the phonons. Indeed, this is the only clear way one can obtain such a separation. 
\end{enumerate}

Given that these \LK{assumptions are readily violated}, what is the reason for their violation?  The answer partly lies in the distinction between
\textit{relative time dynamics} ($\trel=t-t'$) and \textit{average (measurement) time dynamics} ($\tave=\half(t+t')$),
also known as the delay time $\tdelay$.
In equilibrium, the quasiparticle lifetime arises from the decay of
the quasiparticle propagator in relative time;  \LK{through Fourier transformation} this gives rise to a line width.
Once one goes into the measurement time domain, the $\tave$ time axis becomes relevant.  As a function of \textit{average time}, the dynamics
along \textit{relative time} for the Green's function $G$ and the self-energy $\Sigma$ may be different. It is precisely this difference that
gives rise to the (measurement) average time dependence \LK{as will be discussed in the next section.}

This fundamental difference between the quasiparticle lifetime and the measurement time dependence is the primary source
of the \LK{breakdown of the assumptions}. Equilibrium measurements by nature measure in the Fourier domain that comes
from the transform along $\trel$,  for example the equilibrium quantities (self-energy, MDC linewidth) illustrated in
Figs.~\ref{fig:imp_equilibrium} and \ref{fig:ee_equilibrium}.  \LK{Out of equilibrium}, a function
of measurement time these may change in addition to the changes in the quasiparticle distribution.  However,
the dynamics along the average time direction are controlled by energy transfer rather than the dephasing
of the correlation functions along $\trel$, and time domain experiments principally access \LK{the former} aspect of the physics.

\subsection{What determines the population decay rate?}

Given that the self-energy alone does not directly govern the decay rate, the question arises: ``What does?''  
\LK{We can gain some insight into this question by making a connection to the equilibrium situation.}
%
%
For a system at long enough times
after the pump such that $t_\mathrm{delay} \equiv t_\mathrm{ave} -t_\mathrm{pump} \gg \tau_C$ (where $\tau_C$ is
the characteristic dephasing time for the Green's functions and self-energies), the equation of motion for the populations reads
\begin{align}
\frac{d n_\kk(\tave)}{dt_{ave}} \equiv & -i \frac{d G_\kk^<(t,t)}{dt}\bigg\rvert_{t\rightarrow \tave} \\
 = & \int_{-\infty}^t d\bar t \left\lbrace \Sigma^R(t,\bar t) G_\kk^<(\bar t,t) + \Sigma^<(t,\bar t) G^A_\kk(\bar t, t) \right. \nonumber\\
& \left. - G^<_\kk(t,\bar t) \Sigma^A(\bar t, t) - G_\kk^R(t,\bar t) \Sigma^<(\bar t, t) \right\rbrace. \label{eq:rhs_t}
\end{align}
\LK{
The long-time assumption is required for there to be no explicit dependence on the vector potential (which has decayed to
zero by this time).
}


%
If the dynamics along $\tave$ are sufficiently slow that the $G$ and $\Sigma$ are constants as a function of $\tave$ on the
scale of $\tau_C$, we can 
perform a Wigner transformation $(t,t')\rightarrow(\tave,\trel)$ and Fourier transform along $\trel$ to
find
\begin{align}
\frac{d n_\kk(\tave)}{dt_{ave}} = & \int_{-\infty}^\infty d\omega \left\lbrace 
2i \mathrm{Im} \left[ \Sigma^R(\tave, \omega)\right] G_\kk^<(\tave, \omega) 
\right. \nonumber\\
& \left.
 - 2i \mathrm{Im}\left[ G_\kk^R(\tave, \omega)\right] \Sigma^<(\tave, \omega).
 \right\rbrace. \label{eq:rhs_w}
\end{align}
\begin{figure*}[htpb]
	\includegraphics[width=0.99\textwidth,clip=true, trim=0 0 0 0]{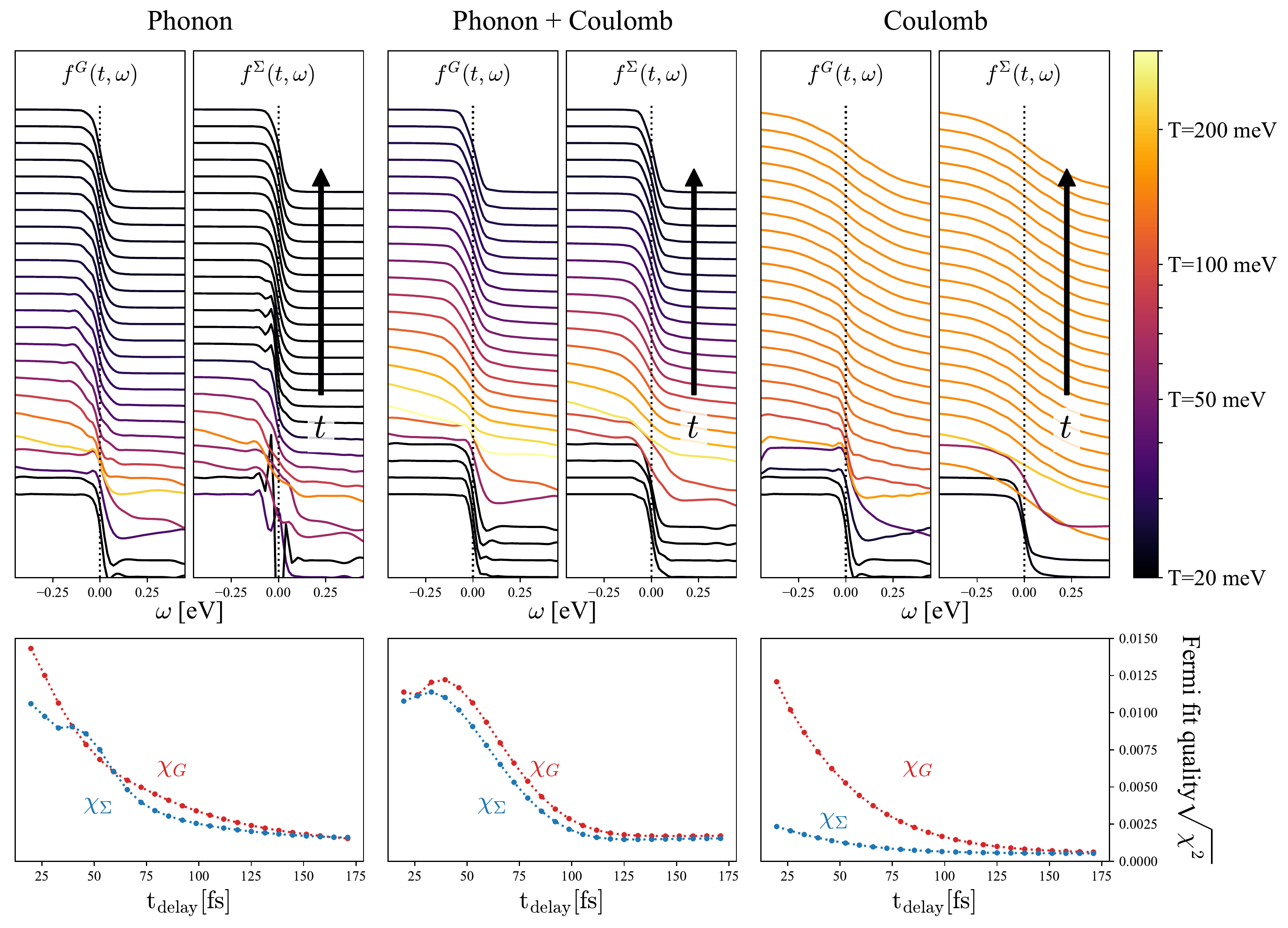}
	\caption{(Color online) Top: Population and interaction distribution functions $f^G(\omega,\tave)$ and $f^\Sigma(\omega,\tave)$
	for three of the cases considered.  The line color indicates the temperature obtained from a Fermi function fit.
	Bottom: $\sqrt{\chi^2}$ corresponding to the Fermi function fits for the three cases above it.}
	\label{fig:FD}
\end{figure*}

\LK{We are now in a position to define the electron and interaction {\it distribution functions}: for the Green's function via
\begin{align}
f^G(\tave,\omega) \equiv i G_\kk^<(\tave,\omega) / 2 \mathrm{Im}\ G_\kk^R(\tave,\omega),
\end{align}
and similarly for the self-energy $\Sigma(\tave, \omega)$.
This definition is motivated by the fluctuation-dissipation theorem, which holds identically in equilibrium with
the distribution functions $f^G(\tave,\omega)=f^\Sigma(\tave,\omega)=n_F(\omega,T)$, where $n_F(\omega, T)$ is
the Fermi distribution function at the system temperature $T$.
}
%
%

Using these distribution functions, we can 
obtain
\begin{align}
\frac{d n_\kk(\tave)}{dt_{ave}} = & |2i|^2\int_{-\infty}^\infty d\omega 
 \mathrm{Im} \left[ \Sigma^R(\tave, \omega)\right] \mathrm{Im} \left[ G_\kk^R(\tave, \omega)\right] \nonumber \\
&\times \left[ f^G(\tave,\omega) - f^\Sigma(\tave,\omega) \right].
\end{align}
For sufficiently well-defined quasiparticles, this can be further simplified to
\begin{align}
\frac{d n_\kk(\tave)}{dt_{ave}} = & |2i|^2 
 \mathrm{Im} \left[ \Sigma^R(\tave, \epsilon_\kk)\right] \mathrm{Im} \left[ G_\kk^R(\tave, \epsilon_\kk)\right] \nonumber \\
&\times \left[ f^G(\tave,\epsilon_\kk) - f^\Sigma(\tave,\epsilon_\kk) \right].
\end{align}

The decay rate of the population at a momentum $\kk$ is thus obtained from the difference in distributions for
the self-energy $\Sigma$ and the populations themselves (through $G_\kk^<$).  From this relation, we can conclude the following:
\begin{enumerate}
\item \LK{We can immediately observe the necessity for allowing different distribution functions for $G$ and $\Sigma$: if they
were equal, then no dynamics would occur because the right hand side would be identically 0.}
\item Although the self-energy is involved, the return to equilibrium is determined by a combination
of $\mathrm{Im}\Sigma^R(\tave,\omega)$ and the distributions.
\item If we use the hot electron model, where the $f^G/f^\Sigma$ functions are thermal functions with the {\it same} temperature,
then there are \textit{no} dynamics
at all except if a different ``temperature'' is assumed for $\Sigma$ and $G$.  In this sense, it is the 
deviations from the hot electron model that arise when $f^G \neq f^\Sigma$ that control the dynamics.
%
\item For (elastic) electron-impurity scattering within the self-consistent Born approximation, the right hand side
of the equation vanishes identically\cite{kemper_relationship_16}.  This is generically the case for \textit{any} self-energy which has the form
\begin{align}
\Sigma^\mathcal{C}  \propto \sum G_\kk^\mathcal{C},
\end{align}
where the sum may be over the entire or a restricted Brillouin zone. This is due to the fact that the distributions $f^G$ and $f^\Sigma$ must be identical in this case.
\end{enumerate}


\LK{The dynamical variable that controls the population dynamics is thus not just the population itself (encoded through $f^G$),
but rather the difference between the distributions for the Green's function $G$ and the self-energy $\Sigma$.  These
distributions $f^G(\tave,\omega)$ and $f^\Sigma(\tave,\omega)$ depends on both frequency and average (measurement)
time. }

\LK{
We have performed the change to Wigner coordinates followed by the Fourier transformation of the full two-time numerical results. 
Fig.~\ref{fig:FD} shows the distribution functions $f^G/f^\Sigma$ for phonon scattering, Coulomb scattering, and the combination as a function
of measurement time and frequency.  The unusual shapes at early times arise due to the Fourier transform overlapping with the
pump for early measurement times.  As time progresses, the sharp step in the distributions (whose origin is the sharp step in the Fermi function at
low temperatures) smoothes out in both the population and interaction distributions.  The subsequent dynamics then depends on
the particular scattering mechanism.  In the case of simple phonon scattering, the distributions quickly return to a low temperature,
and have sharp features due to the phase space restrictions on the phonon scattering.  On the other hand, pure Coulomb scattering
produces smooth, high-temperature-like distributions quickly after the pump.  The mixed case interpolates between the two --
it does return to the low temperature (pre-pump) state, but does so in a smoother way than pure phonon scattering does.}

\begin{figure}[t]
	\includegraphics[width=0.99\columnwidth,clip=true, trim=0 0 0 0]{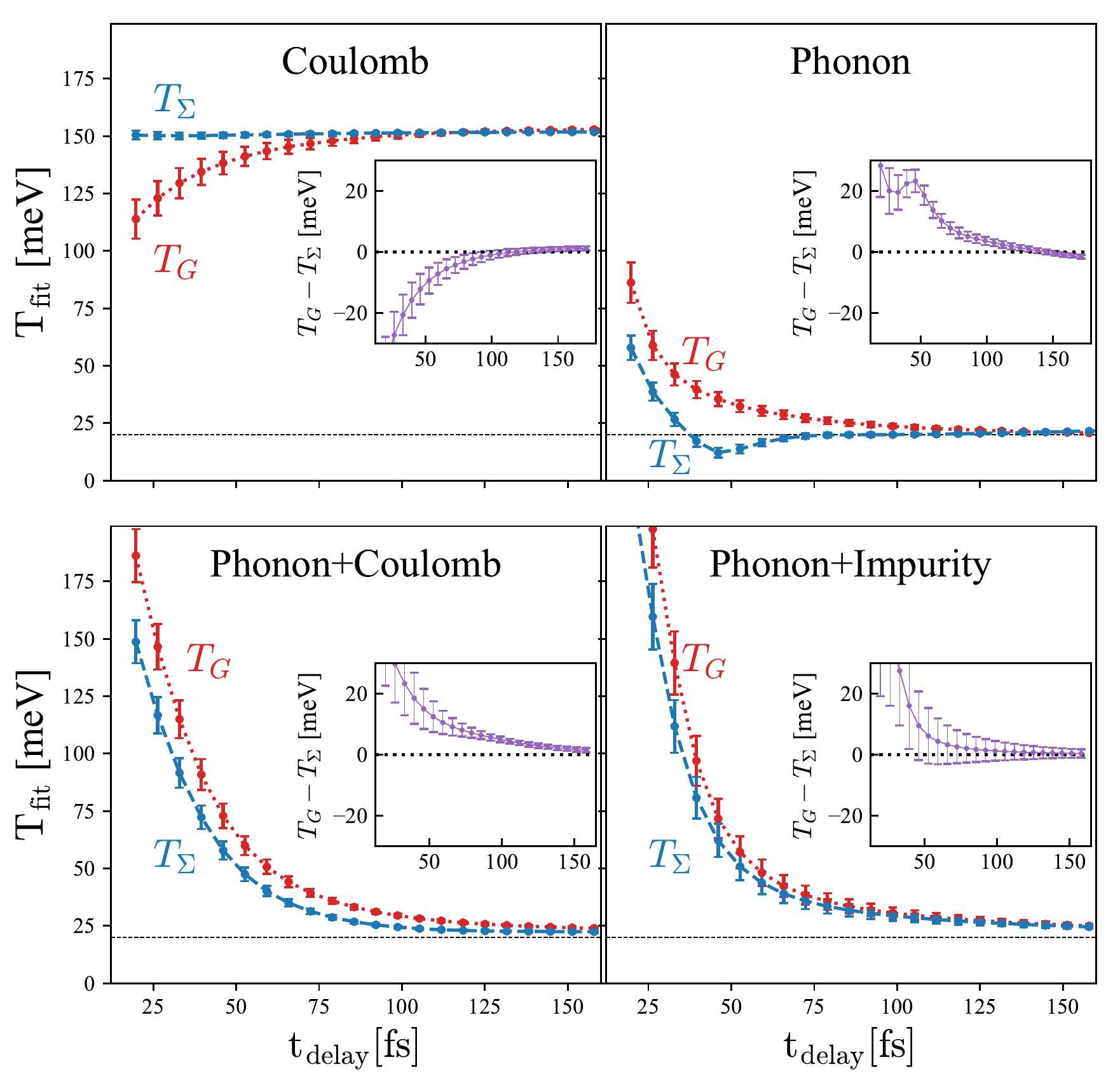}
	\caption{(Color online)  Effective temperatures extracted from fits to Fermi-Dirac functions for the Green's function (populations)
	and self-energy.  The horizontal line indicates the equilibrium temperature before the pump. 
	The insets show the difference between the population and interaction effective temperatures.}
	\label{fig:FDtemp}
\end{figure}

\LK{
As noted above, the dynamics of the populations are determined by the {\it difference} of the population and interaction distribution functions.
To illustrate this, and to make a connection to the concept of quasi-temperature, we have performed Fermi function fits of
the distribution functions $f^G/f^\Sigma$ and plotted the resulting quasi-temperatures $T_G/T_\Sigma$ as a function of time (Fig.~\ref{fig:FDtemp}).
Note that the Fermi function fitting is not perfect, and that there is some deviation from the functional form. This is illustrated
in Fig.~\ref{fig:FD}, where the fit quality (through a $\chi^2$ measure) is shown.  The approach to ``perfect'' Fermi-Dirac form
varies from case to case (and is best for pure Coulomb scattering).  In particular, when phonon scattering is present,
a finite offset to the long-time population remains for long times due to phonon window effects.
The situation with pure impurity scattering is sufficiently far from Fermi functions that fits were not even attempted.}

\LK{
In all cases, the dynamics in Fig.~\ref{fig:FDtemp} 
show that the two fitted quasi-temperatures approach each other at long times.  The case of pure Coulomb scattering is qualitatively
different because of its approach to a high temperature state, as discussed above.  The cases where some phonon
scattering is present all approach the phonon bath temperature, although the pathway varies somewhat
depending on the type of interactions. Since the dynamical variable is not the population distribution function
$f^G$ or the extracted temperature $T_G$, but rather the {\it difference} $(f^G-f^\Sigma)$  or $(T_G-T_\Sigma)$,
we show the temperature difference in the insets.  The difference approaches zero in all cases, but with varying
dynamics, and even with apparently time-dependent rates.
}

In view of these results, it should be noted that the hot electron model is often applied successfully to experimental data\cite{bauer_hot_2015}.  This is not
surprising \textemdash\ the hot electron model typically produces a decaying exponential-like curve, possibly with a final state
that is at some elevated temperature.  This behavior certainly mirrors some of the experimental results,
although it has been previously pointed out that the hot electron model only arises from the limiting case
where electron-electron collisions are faster than the electron-phonon ones\cite{baranov_theory_2014}.
Theoretical work also tends towards thermal solutions because of the typical use of Boltzmann equation
approaches which have thermal states as their fixed point and do not allow for
separate distributions in the population and interactions unless the phonon distribution is explicitly included.
\cite{rethfeld_ultrafast_2002,baranov_theory_2014}

Although this is an exact proof that the assumptions for the hot electron model do not hold in any many-body system,
we cannot easily quantify how close a hot-electron analysis may be to the exact result.  It may in fact be an accurate approximation,
in particular if the long-time dynamics in the self-energy are slow.  However, it is never exact.

\subsection{Implications for experiments}
\LK{
Although the dynamics are determined by a difference between the two
distributions $f^G$ and $f^\Sigma$, typical experiments such as tr-ARPES have ready access to observe only
one of these: $f^G$, as measured by the populations.  As such, it is not immediately possible to determine
the entire right hand side of the equation of motion which would be necessary to extract any information about
the self-energy $\Sigma$.  The question is then: given that only part of the equation ($f^G$) can be observed,
what does its analysis show?}

The answer may be found in 
Figs.~\ref{fig:imprates} and \ref{fig:coulombrates},
where such an analysis was performed.  In discussing these, we noted that although the self-energy may
have multiple components,
the energy-dependent rates are mainly determined by the phonon scattering.  As shown previously,
the contribution from the impurity self-energy typically cancels\cite{kemper_relationship_2016}. 
To understand the remainder after the removal of the impurity scattering within the context of the distribution functions, 
we can rewrite the equation of
motion in terms of the {\it change} in the distributions $\delta f^G/\delta f^\Sigma$:
\begin{align}
\frac{d \left[\delta n_\kk(\tave)\right]}{dt_{ave}}= & |2i|^2\int_{-\infty}^\infty d\omega 
 \mathrm{Im} \left[ \Sigma^R(\tave, \omega)\right] \mathrm{Im} \left[ G_\kk^R(\tave, \omega)\right] \nonumber \\
&\times \left[ \delta f^G(\tave,\omega) - \delta  f^\Sigma(\tave,\omega) \right]. 
\end{align}
where we have subtracted a term proportional to $f^G_\mathrm{eq}(\omega) - f^\Sigma_\mathrm{eq}(\omega) =0$.
This defines
\begin{align}
\delta f^G(\tave,\omega) \equiv & f^G(\tave, \omega) - f^G_\mathrm{eq}(\omega) \\
\delta f^\Sigma(\tave,\omega) \equiv & f^\Sigma(\tave, \omega) - f^\Sigma_\mathrm{eq}(\omega)
\end{align}
since the equilibrium distributions are equal for $G$ and $\Sigma$.
For simple electron-phonon scattering, at high energies and for weak pumping, 
the change in the self-energy $\Sigma(\tave,\omega)$ from equilibrium is typically
not large, and we may approximate $\delta f^\Sigma(\tave,\omega)\approx 0$.  
We can combine the remaining factors through the definition of $f^G$, and for
quasiparticles that are sufficiently well-defined, find
\begin{align}
\frac{d \left[\delta n_\kk(\tave)\right]}{dt_{ave}}
=& 2i \int_{-\infty}^\infty d\omega 
 \mathrm{Im} \left[ \Sigma^R(\tave, \omega)\right] \delta G^<_\kk(\tave,\omega)\\
 \approx & 2i 
 \mathrm{Im} \left[ \Sigma^R(\tave,\epsilon_\kk)\right] \int_{-\infty}^\infty d\omega  \delta G^<_\kk(\tave,\omega) 
 \label{eq:onshell}\\
 =& -2 \mathrm{Im} \left[ \Sigma^R(\tave,\epsilon_\kk)\right]  \delta n_\kk(\tave).
\end{align}
Eq.~\ref{eq:onshell} follows from the well-defined quasiparticle approximation because $G^<_\kk(\omega)$ is sharply
peaked at $\omega=\epsilon_\kk$.

This expression suggests that within this approximation a relatively simple exponential decay may occur, with $2\mathrm{Im }\Sigma^R(\tave,\epsilon_\kk)$
as the time constant. Note that this is different from the hot-electron model, because we have explicitly assumed
a difference between the distribution functions for $G$ and $\Sigma$.

\LK{
This may be extended to any situation where the change in the populations is in a sense larger than
that in the interactions.  Within this limit, the simple picture may also be applied.
The full solution requires the relaxation of the assumptions made in this section, and the result is the modification of the simple
decay picture with the complications discussed in this paper and elsewhere\cite{kemper_effect_2014,rameau_energy_2016}.
This is clearly seen in Fig.~\ref{fig:FDtemp}, where $T_\Sigma(\tave)$ is not negligible.
In addition, inclusion of the phonon degrees of freedom will further make changes to the equations of motion that
will strongly modify $f^\Sigma$, in particular at later times. We reserve a study of this phenomenon for future work.
However, we do know that in the long time limit, the first correction for the case where the time evolution is slow
results in a large number of difference terms that don't seem to have any simple form;
no one term dominated over another\cite{kemper_role_16}. 
}

\LK{
\subsection{When are the assumptions valid?}
As we have discussed throughout this manuscript, there are situations where the assumptions -- although not generally true --
may nevertheless hold.  These may include:
%
%
\begin{itemize}
\item Assumption (i), that many-body systems must relax after excitation, is often not at issue -- empirically, all photoexcited systems
eventually return to their pre-pump equilibrium state.  Our point here is that for the return to equilibrium, there must be an energy reservoir and
energy transfer -- in contrast to a line width which is always present when the self-energy is nonzero.
\item Assumption (ii), that the self-energy governs the decay, is true in limited cases.  In particular, when the interactions are dominated
by a single scattering process between the excited electrons and some bath (e.g. phonons), the decay is determined by the self-energy.
\cite{rameau_energy_2016}.
However, going beyond this simple situation breaks down this assumption.
\item Assumption (iii), that relaxation rates separate in the time domain, holds when each process is associated with a bottleneck
of some sort.  If both scattering processes can happen at the same time, no separation occurs.
\item Assumption (iv), that electrons thermalize among themselves, never holds in the strict sense where both of the 
distributions are determined by the same thermal distribution function.  However, functionally it appears that at least the distribution
function for the populations $f^G$ is often reasonably thermal\cite{bauer_hot_2015}.  As we have shown, however, $f^\Sigma$ cannot be the same 
thermal distribution, or no dynamics would occur.
\end{itemize}
}

\section{Conclusions and outlook}

In this work we have shown by contradiction that several assumptions which are pervasive in non-equilibrium experiments 
\LK{may be violated in certain circumstances.}
To whit, these assumptions are:  
(i) many-body systems relax after excitation; 
(ii) the self-energy governs the relaxation rate; 
(iii) the time domain allows one to separate the relaxation rates from different scattering processes;
(iv) electrons rapidly scattering amongst themselves to create a hot thermal electrons before subsequently scattering with phonons
to obtain a common final temperature.
Although this work is by no means an exhaustive inclusion of all the possible scattering processes, we have presented counter examples 
to each of these \LK{assumptions} through a system of electrons interacting variously with impurities,
phonons, and internally (Coulomb). We have discussed the origin of these \LK{violations} as lying in the difference between relative and
average (measurement) time dynamics.

We have further shown that the hot electron model, which is used widely in the literature, is not to be taken at face value
since a system where a temperature can be defined through the fluctuation-dissipation theorem will not have any dynamics
at all.  Rather, the dynamics are determined by the difference between two distribution functions -- one each for the populations
and one for the interactions.

These observations then raise the question: ``how should one
interpret non-equilibrium experiments, and what leads to the observed dynamics?'' 
The answer, which we have started to address here and in previous work\cite{rameau_energy_2016},
is that the dynamics of energy transfer between various subsystems controls the dynamics, but internal scattering
within the subsystem can cause substantial modifications.
For the situations studied here, a
return to equilibrium only occurred 
when a path (through the phonons) was provided for the excess energy in the electrons to dissipate.
There are similarities between the energy dissipation characteristics and some of the equilibrium quantities since they
share some of the underlying physical principles such as phase space restrictions, leading to e.g. the phonon window effect,
although we should understand this as a phase space restriction on the excess energy leaving the electrons rather than the
single quasiparticle lifetime.  

When it comes to the interpretation of the obtained lifetimes, they should be taken as direct quantitative measures only if it can be shown
that a single process is responsible for the decay and any internal scattering is absent, otherwise the rate may contain many contributions
and it is not clear how to separate these.  Nevertheless, progress may be made by varying the other external control parameters of the
experiment, and observing how the decay rates change as a function of said parameters; this approach is already used in the field, but
should be more strongly considered as a method of investigation using time-resolved techniques.

\begin{acknowledgments}
We would like to acknowledge U. Bovensiepen, T.P. Devereaux, P. Kirchmann and J. Sobota for helpful discussions.
J.K.F. was supported by the U.S. Department of Energy, Office of Basic Energy Sciences, Materials Sciences and Engineering under Contract
No. DE-FG02-08ER46542 and 
also by the McDevitt bequest at Georgetown.
Computational resources were provided by the National Energy Research Scientific Computing Center supported by the U.S. Department of Energy, Office of Science, under Contract No. DE-AC02-05CH11231. 
\end{acknowledgments}

\bibliography{tdrefs}

\end{document}